\documentclass[journal=apchd5,manuscript=article]{achemso}
\usepackage[version=3]{mhchem} % Formula subscripts using \ce{}
\usepackage[utf8]{inputenc}
\usepackage[T1]{fontenc}
\usepackage{color}
\usepackage{soul}
\usepackage{caption}
\usepackage{graphicx}
\usepackage{float}
\usepackage{xr}
\makeatletter
\newcommand*{\addFileDependency}[1]{% argument=file name and extension
  \typeout{(#1)}
  \@addtofilelist{#1}
  \IfFileExists{#1}{}{\typeout{No file #1.}}
}
\makeatother

\newcommand*{\myexternaldocument}[1]{%
    \externaldocument{#1}%
    \addFileDependency{#1.tex}%
    \addFileDependency{#1.aux}%
}

\myexternaldocument{Supp-Inf}

\usepackage{xcolor}
\usepackage[hidelinks,colorlinks=false,citecolor=black,linkcolor=black]{hyperref}

\author{Cynthia Vidal}
\affiliation{The Blackett Laboratory, Department of Physics, Imperial College London, London SW7 2AZ, U.K.}
\email{c.vidal@imperial.ac.uk}
\author{Benjamin Tilmann}
\affiliation{Nano-Institute Munich, Department of Physics, Ludwig-Maximilians-University Munich, 80539 Munich, Germany}
\author{Sunny Tiwari}
\affiliation{Aix Marseille Univ, CNRS, Centrale Marseille, Institut Fresnel, 13013 Marseille, France}
\author{T. V. Raziman}
\affiliation{Department of Mathematics, Imperial College London, London SW7 2AZ, U.K.}
\alsoaffiliation{The Blackett Laboratory, Department of Physics, Imperial College London, London SW7 2AZ, U.K.}
\author{Stefan A. Maier}
\affiliation{School of Physics and Astronomy, Monash University, Clayton, Victoria 3800, Australia}
\alsoaffiliation{The Blackett Laboratory, Department of Physics, Imperial College London, London SW7 2AZ, U.K.}
\author{Jérôme Wenger}
\affiliation{Aix Marseille Univ, CNRS, Centrale Marseille, Institut Fresnel, 13013 Marseille, France}
\author{Riccardo Sapienza}
\affiliation{The Blackett Laboratory, Department of Physics, Imperial College London, London SW7 2AZ, U.K.}
\email{r.sapienza@imperial.ac.uk}
\title{Fluorescence enhancement in topologically optimized gallium phosphide all-dielectric nanoantennas}


\begin{document}
\begin{tocentry}
\includegraphics[width=\linewidth]{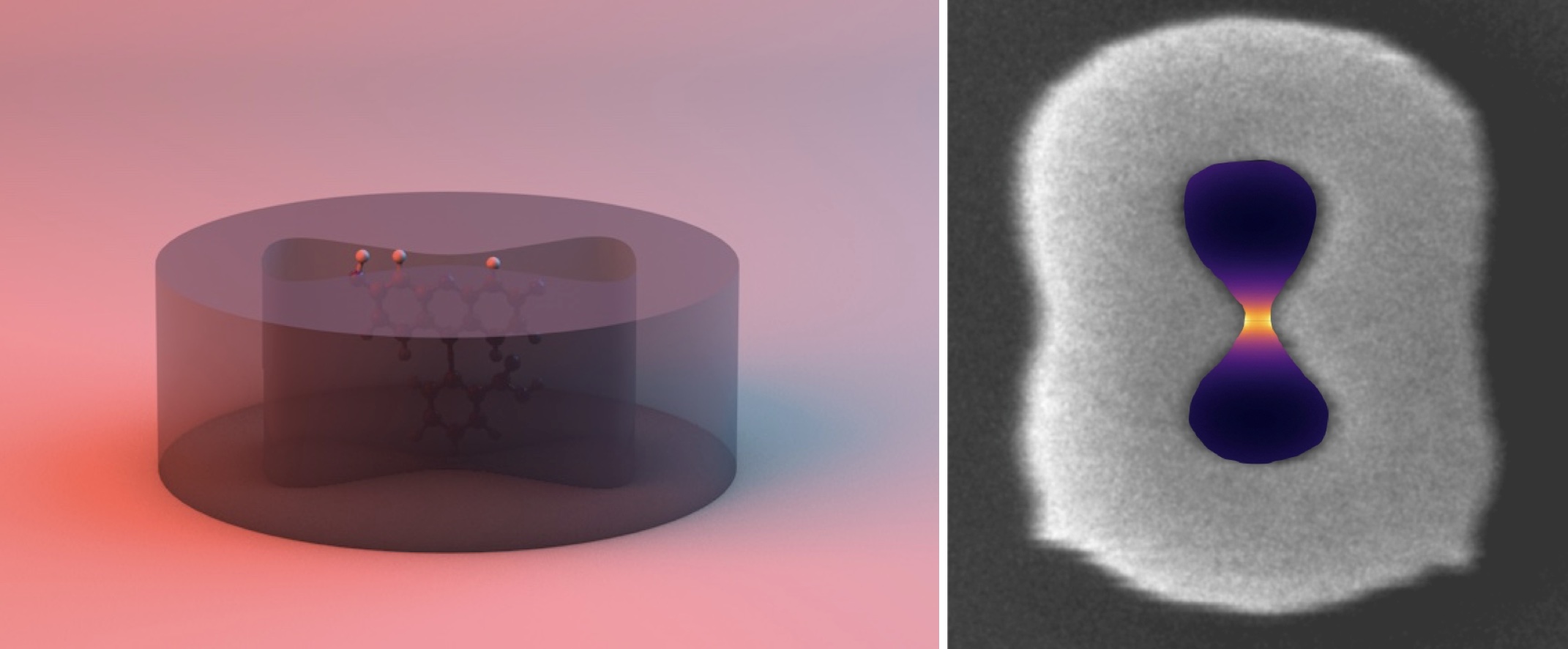}
\end{tocentry}

\begin{abstract}
Nanoantennas capable of large fluorescence enhancement with minimal absorption are crucial for future optical technologies from single-photon sources to biosensing. Efficient dielectric nanoantennas have been designed, however, evaluating their performance at the individual emitter level is challenging due to the complexity of combining high-resolution nanofabrication, spectroscopy and nanoscale positioning of the emitter. Here, we study the fluorescence enhancement in infinity-shaped gallium phosphide (GaP) nanoantennas based on a topologically optimized design. Using fluorescence correlation spectroscopy (FCS), we probe the nanoantennas enhancement factor and observed an  average of 63-fold fluorescence brightness enhancement with a maximum of 93-fold for dye molecules in nanogaps between 20 nm and 50 nm. The experimentally determined fluorescence enhancement of the nanoantennas was confirmed by numerical simulations of the local density of optical states (LDOS). Furthermore, we show that beyond design optimisation of dielectric nanoantennas, increased performances can be achieved via tailoring of nanoantenna fabrication.
\end{abstract}

Light-matter interaction can be controlled by the use of optical nanoantennas which confine light at the nanoscale, resulting in high local fields\cite{novotny_principles_2006}. While plasmonic nanoantennas have long been used for their large Purcell enhancement factor, they suffer from strong absorption and non-radiative quenching losses. Instead, dielectric nanostructures offer moderate Purcell factor combined with close to no losses~\cite{krasnok_all-dielectric_2012,kuznetsov_optically_2016}, by exploiting both electric and magnetic resonances \cite{staude_tailoring_2013,yang_anapole-assisted_2018}. Thus, dielectric nanocavities have recently been the focus of intense research in the field of nanophotonics  \cite{yan_all-dielectric_2020,koshelev_nonradiating_2019} with applications in lasing\cite{koshelev_dielectric_2021}, integrated photonics \cite{diaz-escobar_light_2023}, nonlinear optics \cite{cambiasso_bridging_2017,frizyuk_nonlinear_2021}, and biosensing\cite{ghenuche_matching_2015,cambiasso_surface-enhanced_2018,regmi_all-dielectric_2016}. In particular, gallium phosphide (GaP) presents the advantages of a high refractive index associated with close to zero absorption losses in the visible making it a material of choice to boost fluorescence emission \cite{cambiasso_bridging_2017,sortino_bright_2021}.

Topological optimization stands as a rapidly evolving field aiming to perfect the design of dielectric nanoantennas. The central challenge revolves around maximizing the local density of optical states (LDOS) and intensifying the light-matter interaction at the nanoscale, all achieved through the conduit of all-dielectric antennas \cite{robinson_ultrasmall_2005,liang_formulation_2013,wang_maximizing_2018,molesky_inverse_2018,wu_nanoscale_2021,albrechtsen_nanometer-scale_2022,albrechtsen_two_2022,yang_-chip_nodate}. This avenue not only provides a lossless alternative to plasmonic nanoantennas but also promises heightened performance. The design approaches encompass various strategies, including the utilization of evolutionary algorithms \cite{gondarenko_spontaneous_2006,gondarenko_low_2008,bonod_evolutionary_2019} and the optimization of electric and magnetic dipoles \cite{liang_formulation_2013,hu_design_2016,brule_magnetic_2022}. These efforts have coalesced into the development of generalized bowtie antennas surrounded by reflectors.

In a recent stride forward, our research group has shown that the phase distribution of point-like emitters plays a critical role, even at a deeply subwavelength scale \cite{mignuzzi_nanoscale_2019}. By maximizing the in-phase backscattering into the source dipole, while concurrently mitigating the undermining impact of destructive interference, we have forged a rational architectural framework for all-dielectric antennas. Further improvement using an iterative approach has led to intense electromagnetic LDOS enhancement up to 3 orders of magnitude using a topologically optimized dielectric nanoantenna \cite{mignuzzi_nanoscale_2019}.

Despite the plethora of theoretical insights into the topological optimization of dielectric nanostructures, experimental demonstrations of these hybrid nanoantennas remain sparse and predominantly confined to the near-infrared domain \cite{albrechtsen_nanometer-scale_2022,hu_experimental_2018,hong_anapole-assisted_2023}. This preference stems from the greater ease of fabrication due to the larger wavelength and antenna dimensions. However, in the visible spectrum, the intricacies of nanofabrication and the imperative for precise positioning of the dipole emitter have constrained experimental endeavors \cite{moller_angular_2013,kuzyk_dna_2018,humbert_versatile_2022}. Notably, there has yet to emerge a dielectric nanoantenna that has been both designed and characterized at optical wavelengths utilizing a topologically optimized model. This unexplored area emphasizes the importance of taking significant steps forward to connect theory with real-world applications and further advance nanophotonics.

Here, we bridge this gap and experimentally showcase the performance of gallium phosphide (GaP) nanoantennas designed according to a topologically optimised approach. Our GaP nanoantenna is shaped like the infinity symbol, with a bowtie-shaped nanogap at its center (shown in Figure 1a,b. This design, inspired by the general framework outlined in Mignuzzi's work \cite{mignuzzi_nanoscale_2019}, is strategically crafted to enhance local electromagnetic effects through the precise tuning of constructive interference. Fluorescence correlation spectroscopy (FCS) experiments thoroughly characterize the GaP nanoantennas and assess their optical performance in enhancing single Alexa Fluor 647 molecule fluorescence. Our all-dielectric nanoantennas achieve a remarkable enhancement of the fluorescence brightness up to 90-fold together with optical confinement into a 200 zeptoliter ($10^{-21}$~L) detection volumes, 5000 fold below the confocal diffraction limit. These experimental values stand in excellent agreement with our numerical simulations. This successful experimental demonstration of all-dielectric topologically optimized nanoantennas in the visible spectral range holds profound significance for the realms of future sensing and quantum technologies.

\begin{figure}[ht]
\centering
\includegraphics[width=\linewidth]{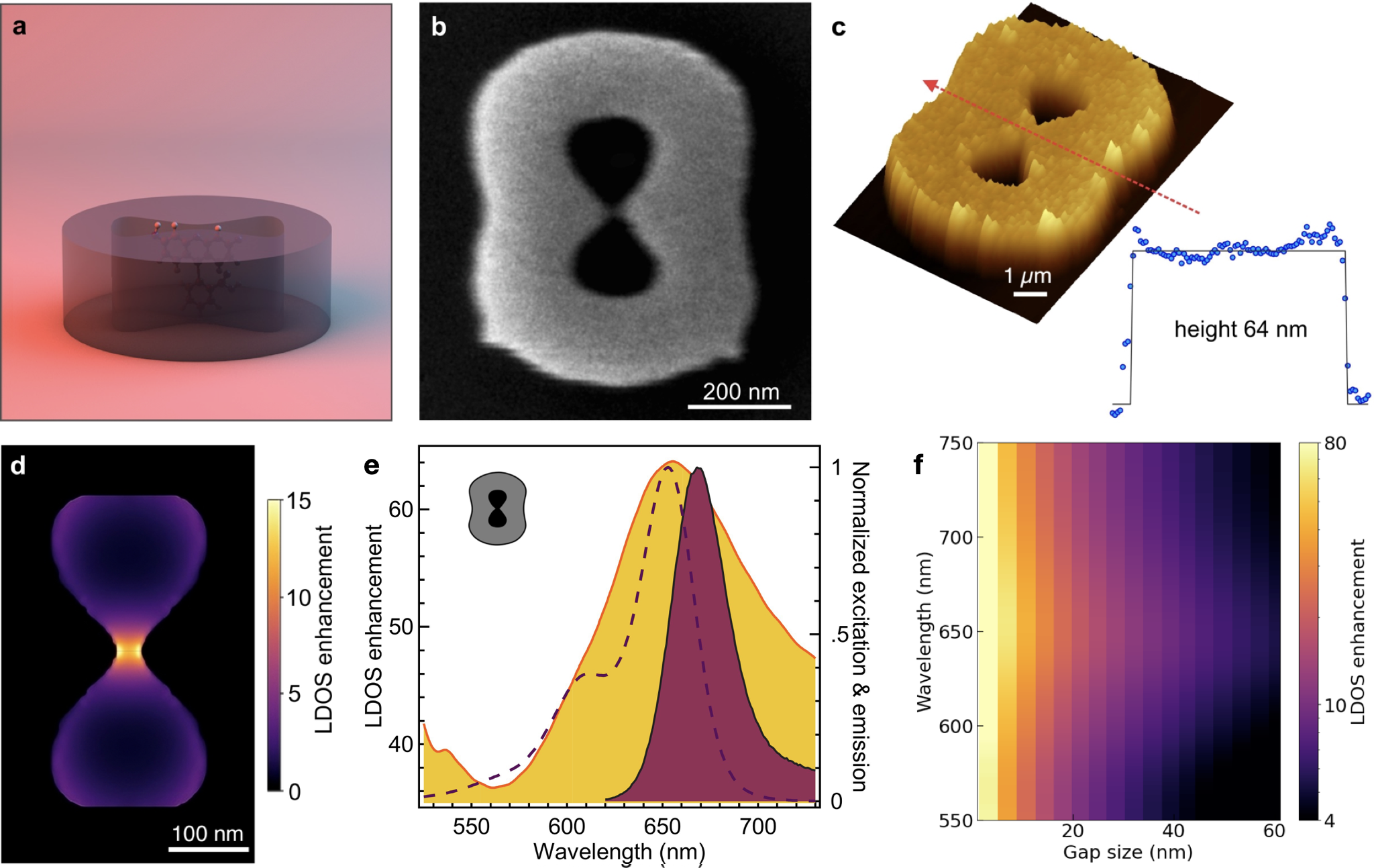}
\caption{(a) All-dielectric topologically optimized nanoantenna to enhance the fluorescence from single diffusing molecules. (b) SEM image of a GaP nanoantenna. (c) AFM measurement of a nanoantenna. (d) Map of the LDOS enhancement at $\lambda = 650$~nm for a 30~nm gap antenna with dipole emitter aligned along the gap. (e) Overlay of the LDOS spectral enhancement (left axis, 6.5 nm gap size) and the fluorescence excitation and emission spectra of Alexa Fluor 647 used in this study (right axis). (f) 2D map of the LDOS enhancement as a function of the gap size and the emission wavelength.}
\label{fig1:nanofab}
\end{figure}

The rationale behind our design departs from the quasi-static approximation to fully consider the phase of the induced polarization currents into the source dipole \cite{mignuzzi_nanoscale_2019}. By enhancing the constructive interference terms and removing the negative influence of destructive interference terms, the local electromagnetic enhancement can by strategically optimized. We fabricated arrays of GaP nanoantennas using standard electron beam lithography (EBL)  followed by reactive-ion etching (RIE). Details and a sketch of the nanofabrication process is illustrated in Figure~\ref{fig_m:nanofab} in Supporting Information . A scanning electron microscope (SEM) image of a typical GaP nanoantenna is shown in Figure \ref{fig1:nanofab}b. 
Nanoantennas were designed with a range of gap sizes from 15~nm to 45~nm, with 660~nm x 510~nm dimensions. The deviations of fabricated antenna dimensions from the design, determined by SEM, are listed in SI Table \ref{T1_dimension}. The GaP film thickness and roughness were determined by AFM measurements on three nanoantennas, yielding heights of $H = 63.9 \pm 0.53$~nm and roughness of approximately $8.75 \pm 0.35$~nm, calculated from RMS after isolating the top structure along the line cut across the middle of the antenna (Figure 1c), and the error is the standard deviation of three measurements.(Figure~\ref{fig1:nanofab}c). Due to proximity effects, the gap is often bridged below 15~± 2.5 nm and is difficult to consistently replicate with a standard EBL setup \cite{xiong_experimental_2021}. 

Numerical simulations predict an intense LDOS enhancement for the topologically optimized GaP antenna, as shown in Figure~\ref{fig1:nanofab}d-f. At resonance, the field is confined in the nanoantenna due to constructive interferences, leading to strong LDOS enhancement in the nanogap region between the two bowtie tips (Figure~\ref{fig1:nanofab}d). In addition, according to Maxwell\textquotesingle s equations, the normal component of the electric displacement field remains continuous at the boundaries between two dielectrics. \cite{robinson_ultrasmall_2005} Therefore, the size of the nanogap being much smaller than the wavelength, a strong, frequency-independent electrostatic enhancement, can be achieved for emitters aligned along the gap direction, so the LDOS enhancement effectively covers a broad spectral range (Figure~\ref{fig1:nanofab}e,f). LDOS enhancement factors exceeding 40-fold are predicted for nanogap sizes below 10~nm (Figure~\ref{fig1:nanofab}f). 

We use fluorescence correlation spectroscopy (FCS) to characterize the enhancement of the fluorescence brightness for a molecule placed in the gap of the GaP nanoantenna. While placing an individual static emitter in the center of the nanogap is highly challenging \cite{glembockyte_dna_2021}, FCS exploits the Brownian motion of the individual molecules diffusing in solution to probe the nanoantenna response \cite{wenger_photonic_2010,wohland_introduction_2020}. FCS consists of measuring the temporal auto-correlation function (ACF) of the fluorescence signal from single molecules in order to determine their brightness. It is an established method which allows the quantification of the radiative enhancement from nanostructures \cite{regmi_all-dielectric_2016,regmi_nanoscale_2015,flauraud_-plane_2017}. 
The fluorescence signal is collected via a confocal microscope and detected by a single-photon counting avalanche photodetector. When the LDOS enhancement from the nanogap occurs, the shape and amplitude of the ACF are modified which in turn allow to estimate the number of molecules and their brightness enhancement within the nanogap volume \cite{regmi_all-dielectric_2016,flauraud_-plane_2017}.

\begin{figure}[htbp]
\centering
\includegraphics[width=\linewidth]{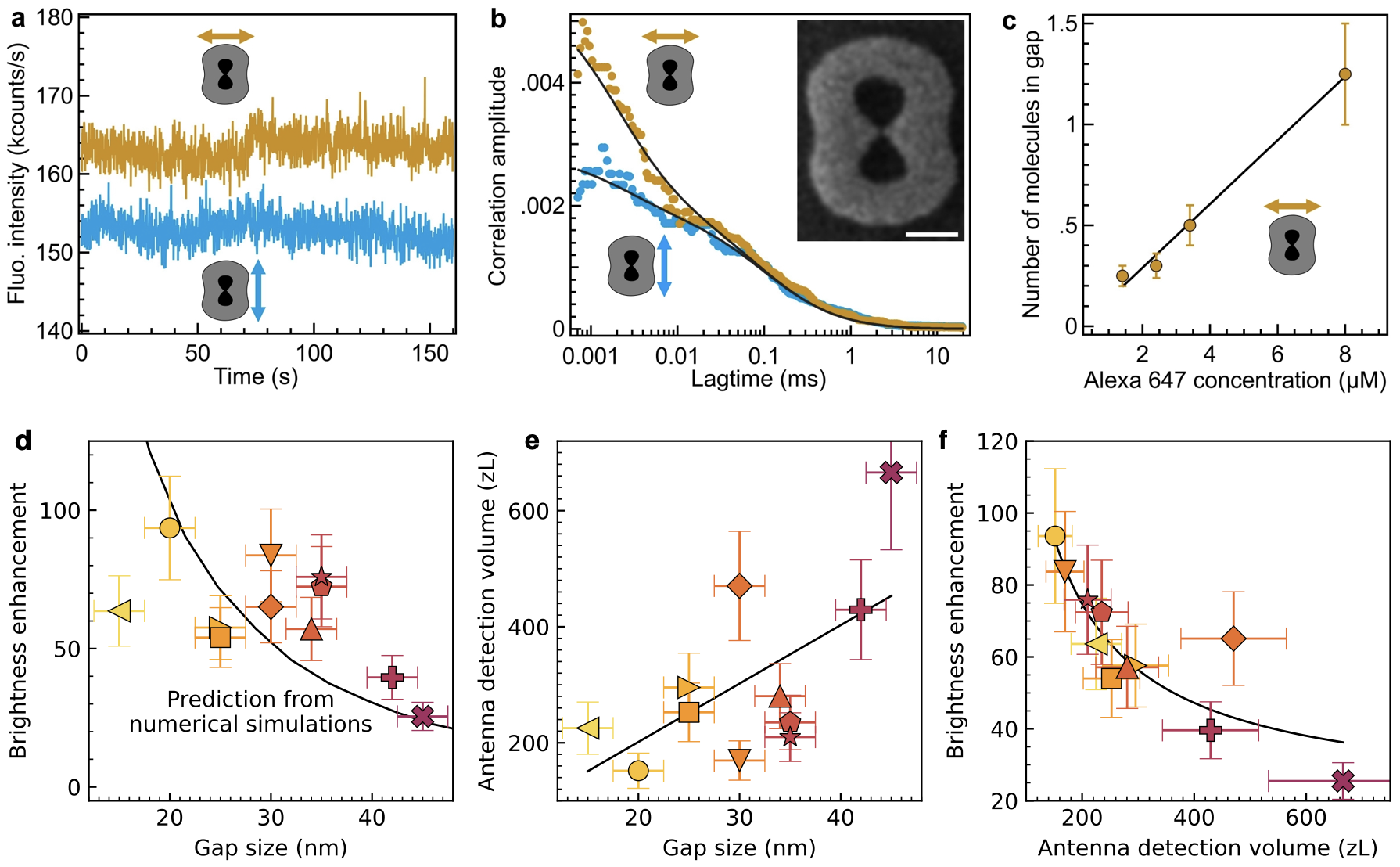}
\caption{Experimental characterization of topologically optimized GaP nanoantennas. (a) Fluorescence intensity time traces recorded on a 1.4 µM solution of Alexa Fluor 647 with 200 mM methylviologen on a GaP nanoantenna with the excitation polarization parallel (orange) or perpendicular (blue) to the 30 nm bowtie nanogap. The binning time is 100 ms. (b) Measured and fitted ACF g($\tau$) as a function of the correlation time $\tau$ from a single GaP nanoantenna with a 30~nm gap imaged in the inset. The arrows indicate the polarization direction of the laser excitation. Scale bar: 200~nm. (c) Evolution of the number of molecules in the nanogap $N^{*}$ as a function of the fluorescent dye concentration. (d) Scatter plot of the fluorescence brightness enhancement as a function of the nanogap size as determined by SEM. Through (d-f), each marker symbol corresponds to a specific GaP nanoantenna. The line is the prediction from numerical simulations, it is not a fit to the experimental data. 
(e) Nanogap detection volume determined by FCS as a function of the nanogap size. The line fit to the data is linear. (f) Fluorescence brightness enhancement as a function of the nanogap detection volume. The line is a fit with a fixed -1 exponent.The error in brightness enhancement measurements is $20\%$ estimated from the FCS experimental uncertainties and the data fit analysis.}
\label{fig2:fcs}
\end{figure}

As a molecular probe, we use Alexa Fluor 647 dyes, with excitation wavelength at 635 nm and emission at 670 nm, where GaP is transparent (Figure~\ref{fig1:nanofab}e). 200~mM of methyl viologen is added to the buffer solution in order to quench the fluorescence quantum yield of Alexa Fluor 647 from 33 to 8\% and increase the magnitude of the fluorescence enhancement factor \cite{punj_plasmonic_2013,puchkova_dna_2015}. This experimental configuration also enables a straightforward comparison with our earlier works using different dielectric and plasmonic antenna designs \cite{regmi_all-dielectric_2016,punj_plasmonic_2013}. 

Figure \ref{fig2:fcs}a,b displays typical experimental results on a 30~nm gap antenna probed with two different excitation polarizations parallel or perpendicular to the nanogap. We have checked that our microscope setup is polarization-insensitive so that the difference seen on the fluorescence intensity time traces (Figure~\ref{fig2:fcs}a) and ACFs (Figure~\ref{fig2:fcs}b) can be directly related to the excitation of the nanogap mode which in turns leads to the fluorescence enhancement. We apply a similar analysis of the FCS fit as in our earlier studies \cite{regmi_all-dielectric_2016,punj_plasmonic_2013} to extract for each antenna the average number of molecules $N^{*}$ in the effective volume defined by the nanogap together with the average fluorescence brightness per emitter $Q^{*}$. From the knowledge of $N^{*}$ and the fluorescent dye concentration, we can then compute the effective detection volume of the nanogap region. The fluorescence brightness enhancement is obtained by dividing the brightness per emitter in the nanogap $Q^{*}$ by the reference brightness per molecule $Q_0$ found with the diffraction-limited confocal configuration. All the fit results for the data in Figure~\ref{fig2:fcs}a,b are summarized in the SI Table \ref{T2_FCS-fit}. We find a linear dependence between the number of molecules measured in the nanogap  $N^{*}$ and the Alexa 647 concentration used in the experiments (Figure~\ref{fig2:fcs}c). This provides an important confirmation of the validity of our results and demonstrates a good reusability of our GaP nanoantennas.

Ideally one would like to directly record the fluorescence lifetime reduction and Purcell enhancement on each GaP nanoantenna. However, this is not currently possible in our setup for two main reasons. First, because we rely on diffusing molecules to probe the nanogap, we have to work at high micromolar concentrations and thus there is a significant number (about 500) of molecules diffusing away from the antenna hot spot but still present in the diffraction-limited confocal volume. As a result of these non-enhanced molecules contribution, there is a significant non-fluctuating fluorescence background overlaid on the antenna hot spot signal. The second reason is that to make the hot spot contribution more apparent in the FCS functions and maximize the brightness enhancement, we use low quantum yield emitters. These molecules have a short fluorescence lifetime around 380~ps~\cite{regmi_all-dielectric_2016} which is below the 600~ps resolution of our current instrument. Further accelerating the decay dynamics with the Purcell enhancement in the nanoantenna leads to a fluorescence lifetime totally beyond the capabilities of our system. This is why we rely on FCS to assess the antenna performance and cannot use Time Correlated Single Photon Counting.

For each individual nanoantenna, we correlate the gap size determined by SEM with the measured brightness enhancement (Figure~\ref{fig2:fcs}d). Our results show a clear increase in brightness enhancement for smaller gap sizes consistent with the enhancement stemming from the hotspot in the nanogaps. Enhancement factors exceeding 60-fold are readily observed for nanogaps below 30~nm. To support our findings, we simulated the brightness enhancement using Lumerical for a dipole emitter with 8\% quantum yield aligned in the centre of the nanogap and an excitation intensity well below the saturation as in the experiment (see Methods). The solid line in Figure~\ref{fig2:fcs}d deduced from the numerical simulations without any free parameter shows a remarkable agreement with the experimental data.

Along with the brightness enhancement per molecule, our FCS measurements simultaneously monitor the evolution of the nanoantenna detection volume with the gap size (Figure~\ref{fig2:fcs}e). Detection volumes below 300 zL  are achieved with nanogaps below 30~nm.  By integrating the Purcell factor $P$ over the gap volume $Hdxdy$, $\frac{\int P(x,y)*Hdxdy}{\max [P(x,y)]}$, in the numerical simulations in Figure~\ref{fig1:nanofab}d, we can estimate a detection volume around 400~zL which comes close to the $215 \pm 50$~zL measured for 30~nm gap antennas with FCS. The nanogap volume scales linearly with the gap size (Figure~\ref{fig2:fcs}e) while the LDOS enhancement in the gap decays inversely proportional to the gap size as seen from simulations (Figure~\ref{fig2:fcs}d). Since the brightness enhancement is proportional to the square of LDOS enhancement for low-quantum yield emitters, we expect the brightness enhancement to decay slightly sublinearly with the mode volume when accounting for the background enhancement contribution from the antenna. This observed correlation echoes findings from previous studies on gold dimer nanoantennas \cite{flauraud_-plane_2017}, reaffirming the credibility of our results.

The nanoantenna with a 15 nm gap does not follow this trend. This may be the consequence of fabrication imperfections such as reduced hotspot efficiency due to a non-smooth nanogap, or high losses due to an increase in the imaginary part of the refractive index. \cite{albrechtsen_nanometer-scale_2022}. Alternatively, individual molecules might be unable to access the centre of the smaller nanogap due to reduced Brownian motion and/or blockage from non-fluorescent buffer molecules adsorbed onto the GaP. However, as the only antenna with such a small gap, we must be careful about generalising the deteriorated performance.

\begin{figure}[ht]
\centering
\includegraphics[width=\linewidth]{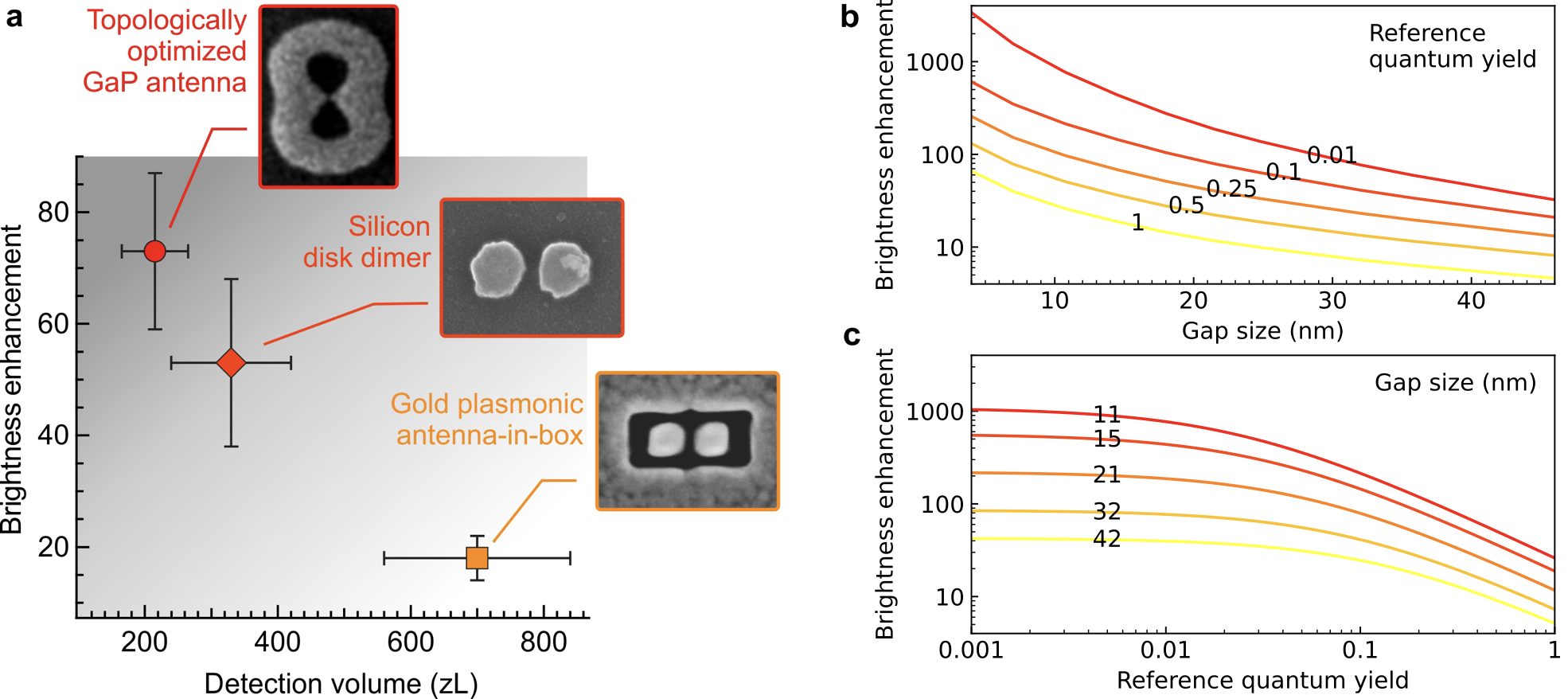}
\caption{Comparison and performance assessment of topologically optimized GaP nanoantennas. (a) 2D map of the fluorescence enhancement and detection volume comparing different optical nanoantennas: the topologically optimized GaP antenna (this work), the silicon disk dimer\cite{regmi_all-dielectric_2016} and the gold plasmonic antenna-in-box \cite{punj_plasmonic_2013}. Importantly for the comparison, the nanoantennas indicated here all share a similar 30~nm gap and were probed using the same fluorescent dye (Alexa Fluor 647 with 200~mM methyl viologen in the buffer). 
(b,c) Numerical predictions of the fluorescence enhancement as a function of the nanogap size and the reference quantum yield of the emitter in homogeneous environment. In (b) the different numbers associated to each curve indicate the initial quantum yield of the emitter considered for the simulations while in (c) the numbers denote the GaP antenna gap size.}
\label{fig3:comp}
\end{figure}

The optical performance of these topologically optimized GaP nanoantennas significantly outperforms the values achieved using silicon nanodisk dimers \cite{regmi_all-dielectric_2016} or gold antenna-in-box \cite{punj_plasmonic_2013} with similar gap sizes.
To ensure a fair comparison, we focus on nanoantennas with similar 30~nm gap sizes probed under similar experimental conditions with the same fluorescent dye. Figure~\ref{fig3:comp}a shows a bi-dimensional map allowing to compare at a glance between the brightness enhancement and the detection volume achieved with different nanogap antennas. Importantly, our topologically optimized nanoantenna outperforms its competitors on both the brightness enhancement and the optical confinement, demonstrating the superiority of its rational phase optimization design \cite{mignuzzi_nanoscale_2019}.

The excellent agreement found between the experimental data and the numerical simulations in Figure~\ref{fig2:fcs}d allows us to elaborate on the simulations to predict the conditions leading to maximum brightness enhancement. The results summarized in Figure~\ref{fig3:comp}b,c predict enhancement factors exceeding 1000-fold for all-dielectric GaP nanoantennas, albeit for emitters with quantum yields below 2\% and gap sizes below 10~nm. This positive insight holds promising implications for the realm of all-dielectric nanophotonics, providing an added incentive to enhance nanofabrication technology and attain sub-10 nm gaps. A narrowing of the nanogap is clearly one of the best ways to improve a nanoantenna’s performance, however it remains extremely challenging to consistently control on  multiple structures \cite{chengfeng_nanofabrication_2023,lyon_gap_2013,manfrinato_resolution_2013}.

In conclusion, we have successfully demonstrated the superior optical performance of all-dielectric GaP nanoantennas designed according to a topologically optimised approach. Thanks to a precise tuning of interferences occurring in the near field of the emitter \cite{mignuzzi_nanoscale_2019}, the LDOS enhancement is maximized, leading to intense brightness enhancement of single quantum emitters. Nanoantennas capable of large fluorescence enhancement with minimal absorption losses are key elements to advance optical technologies from single-photon sources to biosensing. Therefore, this experimental demonstration of all-dielectric topologically optimized nanoantennas in the visible spectral range holds profound significance for the realms of future sensing and quantum technologies. Beyond the design optimisation using a rational approach, increased antenna performances can be achieved via tailoring the nanofabrication to reach smaller gaps. Localising the emitter into the nanogap, controlling its orientation or using new high refractive index materials like transition metal dichalcogenides are other future optimization directions.

\section{Experimental}
\subsection*{Nanofabrication}
Here, we base the nanoantenna design on the one from Ref. \cite{mignuzzi_nanoscale_2019} i.e. 50 nm thick, 550 nm long and 10 nm gap size. For this, a GaP film is deposited on a glass substrate using sputter deposition at $350\,^\circ\text{C}$. Next, the nanofabrication is carried out using EBL and subsequent RIE, as sketched in Figure \ref{fig_m:nanofab} in the Supporting Information. Poly(methyl methacrylate) (PMMA) is used as photoresist and the EBL process is carried out at an acceleration voltage of 30 kV and an aperture size of $30\,\mu m$. Afterwards, the development is done by rinsing the sample in a mixture of methyl isobutyl ketone and isopropyl alcohol (ratio 1:3) for 45 seconds. The gold mask is then deposited using electron-beam evaporation at ultra-high vacuum, with a dedicated thickness of 40 nm.  After the lift-off in an acetone bath, where the remaining PMMA is dissolved, the designed structures remain as gold mask on top of the GaP film. Finally, the structures are transferred into the GaP by performing inductively-coupled plasma RIE based on chlorine gases, after which the remaining gold is removed by respective wet chemistry. 

\subsection*{Optical microscopy experiments}
The FCS measurements are performed using a custom-built confocal microscope (Nikon Ti-U Eclipse) equipped with a water immersion objective (Zeiss C-Apochromat 63x, 1.2 NA). A focused linearly-polarized pulsed laser at 635 nm, with 70 ps pulse duration and 40 MHz repetition rate (LDH series laser diode, PicoQuant) illuminates individual nanoantennas. The antenna sample is immersed in a buffer solution of Alexa Fluor 647 at micromolar concentration with 200~mM methyl viologen as a quencher and 10~mM glutathione as an antioxidant and photostabilizer. Methyl viologen is used to improve the antenna apparent brightness enhancement and make the FCS signature from the nanogap stand out more clearly. With this 200~mM methyl viologen concentration, the quantum yield of the Alexa Fluor 647 dyes is quenched down to $8\%$ \cite{punj_plasmonic_2013,regmi_all-dielectric_2016}. The fluorescence emission in the 650 to 690 nm range is collected by the same microscope objective in the epifluorescence mode. A multiband dichroic mirror (ZT 405/488/561/640rpc, Chroma) and emission filters (ZET405/488/565/640mv2 and ET655 from Chroma plus one FF01-676/37 from Semrock) reject the baskscattered laser light. Detection is performed with a single-photon counting avalanche photodiode (Perkin Elmer SPCM AQR 13) whose output is connected to a time-correlated single photon counting module (HydraHarp 400, Picoquant).
Throughout all our experiments, the laser power measured at the microscope back entrance is kept constant at 2 µW and the total integration time per FCS experiment is 240~s. To efficiently remove the afterpulsing artefacts in the ACFs, we implement the FLCS correction following the approach in Ref.\citenum{enderlein_using_2005} and the built-in function in Symphotime64 (Picoquant).

\subsection*{Fluorescence Correlation Spectroscopy analysis}
FCS computes the temporal correlation of the fluorescence signal $\langle I(t).I(t+\tau) \rangle / \langle I(t) \rangle ^2$, where $\tau$ is the delay (lag) time, and $\langle \, \rangle$ indicates time averaging. Our analysis approach builds on the similar methodology used for our earlier studies on plasmonic \cite{punj_plasmonic_2013,regmi_nanoscale_2015,flauraud_-plane_2017} and dielectric nanoantennas \cite{regmi_all-dielectric_2016}. The total fluorescence signal is considered to be composed to two parts: the enhanced fluorescence from molecules within the nanogap and the fluorescence from the molecules away from the nanogap yet still present within diffraction-limited confocal volume. An essential feature in FCS is that the molecules contribute to $G$ in proportion to the square of their fluorescence brightness, so that the fluorescence from molecules in the nanogap region experiencing the maximum enhancement will have a major contribution in the FCS correlation. See SI for details on the fit analysis. 

\subsection*{Numerical simulations}
We performed electrodynamic simulations using Lumerical, a commercial finite difference time domain (FDTD) solver \cite{noauthor_ansys_nodate}. For details on the numerical simulations, see SI.

\begin{acknowledgement}
This work has received funding from the European Union’s Horizon 2020 research and innovation programme under the Marie Skłodowska-Curie grant agreement No.~882135-BrightNano-vdW and the European Research Council (ERC) grant agreement 723241.
This work was funded by EXC 2089/1-390776260. The authors also acknowledge the support of EPSRC (EP/T027258/1 and EP/P033431/1). 
S.A.M. additionally acknowledges the LeeLucas Chair in Physics and the Centre of Excellence in Future Low-Energy Electronics Technologies, Australian Research Council (CE170100039).
\end{acknowledgement}

\begin{suppinfo}

Additional information on nanofabrication, FCS fit and numerical simulations can be found in the Supporting Information: 

\end{suppinfo}

\bibliography{Main}
\end{document}

\flushbottom
\maketitle

\thispagestyle{empty}
\newpage
\section*{Methods}
\subsection*{Sketch of the nanofabrication procedure}

\begin{figure}[htbp]
\centering
\includegraphics[width=\linewidth]{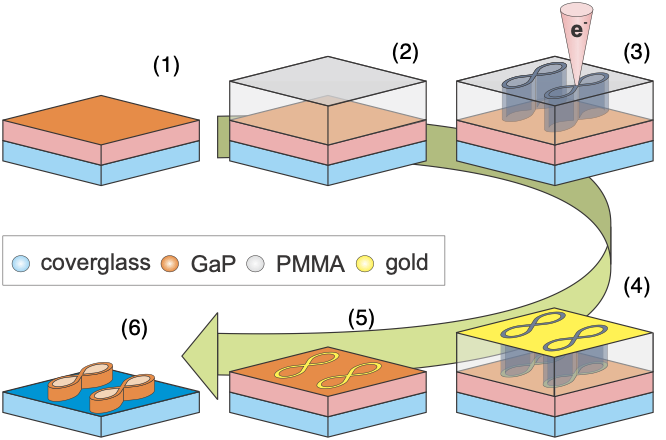}
\caption{First a 50~nm film of gallium phosphide (GaP) is sputtered onto a glass coverslip (1). A layer of poly(methyl methacrylate) (PMMA) is spin-coated on the GaP film (2) before EBL of the infinity antenna structures is performed (3). The sample is then developed to remove the exposed PMMA and coated with a 40-50~nm thick layer of gold (Au) (4). The unexposed PMMA+Au layer is removed via chemical liftoff thus leaving Au only on the structured areas of GaP (5). Reactive ion etching removes the GaP left unshielded by the Au and a final chemical etching removes the last layer of gold leaving only the structured GaP infinity antenna on the coverslip (6).}
\label{fig_m:nanofab}
\end{figure}

\begin{table}[htbp]
\centering
\begin{tabular}{ |p{3cm}|p{3cm}|p{3cm}|p{3cm}|  }
 \hline
 Length (nm)&	Width (nm)&	Gap (nm)\\
 \hline
620   & 490    &45\\
500 &   390  & 35\\
690 & 540 & 25\\
550    &435 & 25\\
745&   585& 30\\
740 & 565& 25\\
740& 590& 35\\
695& 535& 15\\	
615&	475& 35\\
765&	585& 35\\
550&	430& 30\\
 \hline
 \end{tabular}
 \caption{Dimensions of GaP nanoantennas determined from SEM images. Length and width measured at broadest points of the nanoantennas.}
\label{T1_dimension}
\end{table}

\newpage
\subsection*{AFM profile of a gapped nanoantenna}

\begin{figure}[htbp]
\centering
\includegraphics[width=.5\linewidth]{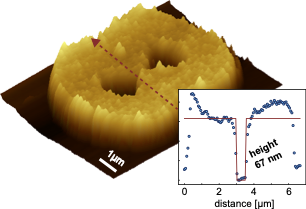}
\caption{3D AFM profile of a gapped nanoantenna. Line cut across the center showing a $40$~nm gap, $67$~nm height}
\label{fig_m:AFM}
\end{figure}

\newpage
\subsection*{Fluorescence Correlation Spectroscopy: Fit analysis}
 The temporal correlation of the fluorescence intensity can be written as:
\begin{equation}\label{Eq:FCS1}
   G(\tau) = \frac{N^* Q^{*2} G_d^*(\tau) + N_0 Q_0^2 G_{d0}(\tau)}{(N^* Q^* + N_0 Q_0)^2} = \rho_1  G_d^*(\tau) + \rho_2 G_{d0}
 \end{equation}
\noindent where $N^*$ is the number of molecules within the gap region with brightness $Q^*$, and $N_0$ is the number of molecules with brightness $Q_0$ diffusing away from the region of interest. $\rho_1$ and $\rho_2$ are the amplitude of a 2-species FCS fit while $G_d^*(\tau)$ and $G_{d0}(\tau)$ are the normalized correlation functions for each species taken individually based on a classical three dimensional model:
\begin{equation}\label{Eq:FCS2}
   G_{di}(\tau) =  \frac{1}{(1+\tau/\tau_{d,i})\sqrt{1+s_i^2 \, \tau/\tau_{d,i}}}
 \end{equation}
$\tau_{d,i}$ stands for the mean residence time (set by translational diffusion) and $s_i$ is the ratio of transversal to axial dimensions of the analysis volume, whose value is set to $s=0.2$ and has negligible influence on the estimates of the number of molecules and brightness within the gap ($N^*$, $Q^*$).

The number of molecules within the gap $N^*$ and their fluorescence brightness $Q^*$ are extracted from the 2-species FCS fit amplitudes $\rho_1$ and $\rho_2$, with the additional knowledge of the total fluorescence intensity $I = N_0 Q_0 + N^* Q^*$ directly measured by our instrument:
\begin{equation}
N^* = \frac{(I- N_0 Q_0)^2}{I^2(\rho_1+\rho_2)-N_0 Q_0^2}
\label{eq:fcsSpie4}
\end{equation}
\begin{equation}
Q^* = \frac{I^2(\rho_1+\rho_2)-N_0 Q_0^2}{(I- N_0 Q_0)}
\label{eq:fcsSpie3}
\end{equation}

The last step to compute $N^*$ and $Q^*$ is to estimate the number of molecules $N_0$ and brightness $Q_0$ for the molecules diffusing away from the nanogap hot spot. To this end, we use the FCS results recorded on the same nanoantenna when the excitation polarization is rotated by $90^{\circ}$ to be perpendicular to the dimer axis.
As additional control, the fluorescence brightness $Q_0$ found with perpendicular polarization is similar to the value found for the confocal reference, while the number of molecule $N_0$ diffusing away from the hot spot is approximately half that seen in the diffraction-limited confocal volume. We relate this effect to the presence of the glass coverslip interface located at the laser focus which cuts the confocal detection volume by a factor of 2.

\newpage
\subsection*{Fluorescence Correlation Spectroscopy: Table of fit values}

\begin{table}[htbp]
\centering
\begin{tabular}{ |p{3cm}||p{3cm}|p{3cm}|p{3cm}|  }
 \hline
 Polarization&	Parallel&	Perpendicular\\
 \hline
I (kcounts/s)   & 162.9    &147.1\\
$\rho_{1} (10^{-3})$ &   3.8  & 0.3\\
$\rho_{2} (10^{-3})$ & 2.0 & 2.1\\
$N_{0}$    &500 &\\
$Q_{0}$ (kcounts/s)&   0.32&\\
$N^{*}$ & 0.2&\\
Volume (zL)& 235&\\
$Q^{*}$ (kcounts/s)& 22.9&\\	
Enhancement&	72.4&\\
 \hline
 \end{tabular}
 \caption{List of fitting parameters for FCS fits in Figure 2a,b. I is the fluorescence intensity, $\rho_{1}$ and $\rho_{2}$ are the values of the ACF at zero delay time considering a model with two diffusion times.}
 \label{T2_FCS-fit}
\end{table}

\newpage
\subsection*{Numerical simulations}

\subsubsection{Model}

The geometric structure of the antenna used in the simulation was created by importing the two-dimensional cross-section from a scanning electron micrograph (SEM) and raising it to a height of 50 nm. Antennas with different gap sizes were created by distorting the central region of the SEM to increase the gap size without affecting the other dimensions.
We have considered an overall average background refractive index $n_{b}= 1.41$ to account for both the glass substrate and surrounding aqueous solution.
The dielectric function of GaP was taken from Ref. \citenum{aspnes_dielectric_1983}.

The brightness enhancement in the nanoantenna results from three processes\cite{regmi_all-dielectric_2016}: (1) excitation enhancement due to the concentration of the electric field, (2) quantum yield enhancement due to the Purcell effect, and (3) increased collection efficiency due to antenna beaming effect. These three processes can be disentangled through numerical simulations and accounted for individually. Here, as the wavelength of interest is sufficiently above the band gap, we neglected the ohmic losses in the antenna.

The enhancement factors of local density of optical states (LDOS), for different combinations of locations of the emitter, were evaluated by performing separate simulations with an electric dipole source for each combination, and obtaining the Purcell factor $P$.

We computed the total fluorescent enhancement $E$ of the antenna by combining the excitation intensity enhancement $F$, the LDOS enhancement $P$, and the collection efficiency modification $\gamma$, following the method in Ref.\citenum{kern_molecule-dependent_2012}
\begin{align}
    E = \displaystyle\frac{\gamma \ P}{I \sigma / k_0 + \left[(P-1)\eta_0 + 1\right] / F} \left(1 + \frac{I \sigma}{k_0}\right) \,
\end{align}
where $k_0$ is the intrinsic total decay rate of the molecule with intrinsic quantum yield $\eta_0$, $\sigma$ is the absorption cross section, and $I$ is the illumination intensity.For this calculation, the emitter is placed at the centre of the gap, aligned along the gap.

As our measurements were undertaken at a laser fluence well below saturation to guarantee a linear dependence with the laser power, we used the simplification of the low-intensity regime
\begin{align}
    E \approx \displaystyle\frac{\gamma~P~F}{1-\eta_0 + \eta_0~P} \,.
\end{align}
\newpage
\subsubsection{Contribution of the incident field}
\begin{figure}[ht!]
\centering
\includegraphics[width=\linewidth]{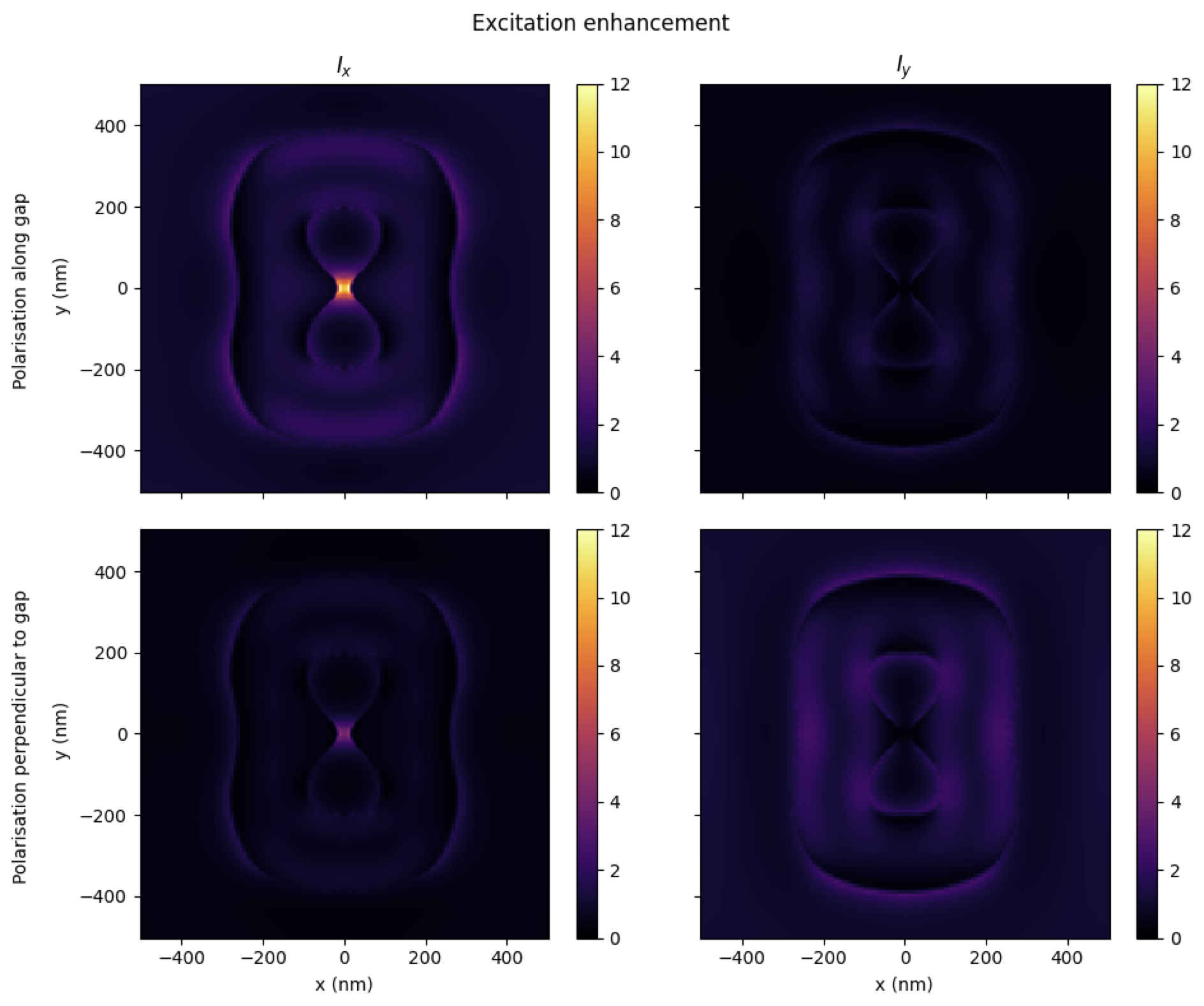}
\caption{The incident field enhancements, $|E|/|E_{0}|$, for both incident polarization computed by averaging local enhancements over multiple illuminations within the numerical aperture and treating the polarised intensities according to chapter 3 in Ref. \citenum{novotny_principles_2006}}
\label{fig_m:excitation}
\end{figure}

\newpage
\subsubsection{Spectral mode decomposition}

\begin{figure}[ht!]
\centering
\includegraphics[width=\linewidth]{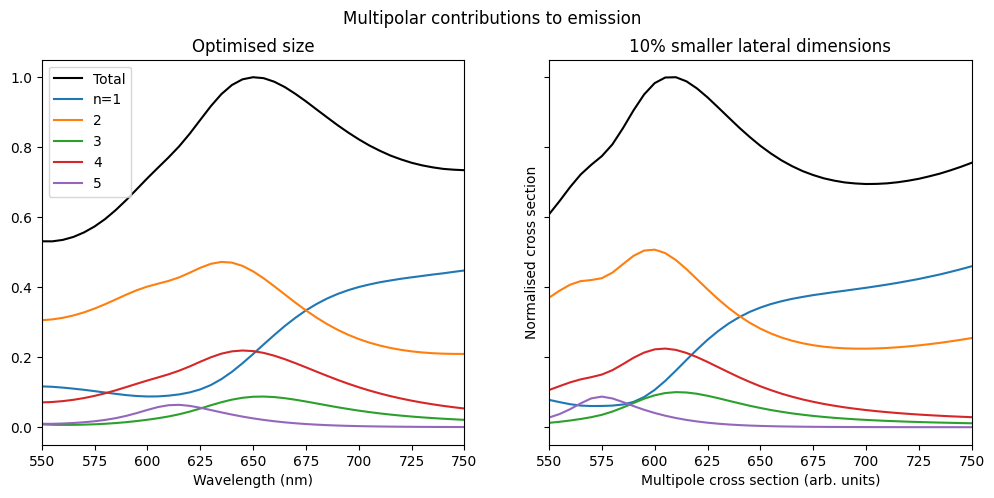}
\caption{Spectral responses of a nanoantenna with an optimal size (left) and one with a size corresponding to the fabricated ones, i.e. with $10\%$ smaller lateral dimensions (right).}
\label{fig_m:multipoles}
\end{figure}

\bibliography{Main}